\documentclass[pra,aps,showpacs,twocolumn]{revtex4}

\usepackage{epsfig}
\usepackage{epsf}
\usepackage{bm}                 
\usepackage{amsmath}
\usepackage{amssymb}
\usepackage{dcolumn}
\usepackage{graphicx}

\def\unity{\mbox{\small 1} \!\! \mbox{1}}
\def\real{\mbox{\rm Re}}
\def\imaginary{\mbox{\rm Im}}

\begin{document}

\title{Gravitational decoherence}

\author{Pieter Kok} \email{Pieter.Kok@jpl.nasa.gov}

\affiliation{Quantum Computing Technologies Group, Jet Propulsion 
Laboratory, California Institute of Technology \\ Mail Stop 126-347, 
4800 Oak Grove Drive, Pasadena, California 91109-8099}

\author{Ulvi Yurtsever} \email{Ulvi.Yurtsever@jpl.nasa.gov}

\affiliation{Quantum Computing Technologies Group, Jet Propulsion 
Laboratory, California Institute of Technology \\ Mail Stop 126-347, 
4800 Oak Grove Drive, Pasadena, California 91109-8099}

\date{\today}

\begin{abstract}
 We investigate the effect of quantum metric fluctuations on qubits
 that are
 gravitationally coupled to a background spacetime. In our first example,
 we study the propagation of a qubit in flat spacetime whose metric is
 subject to flat quantum fluctuations with a Gaussian spectrum. We find
 that these fluctuations cause two changes in the state of the qubit:
 they lead to a phase drift, as well as the expected exponential
 suppression (decoherence) of the off-diagonal terms in the density
 matrix. Secondly, we calculate the decoherence of a
 qubit in a circular orbit around a Schwarzschild black hole. The
 no-hair theorems suggest a quantum state for the metric in which the
 black hole's mass fluctuates with a thermal spectrum at the Hawking
 temperature. Again, we find that the orbiting qubit undergoes
 decoherence and a phase drift that both depend on the temperature of
 the black hole. Thirdly, we study the interaction of coherent and
 squeezed gravitational waves with a qubit in uniform motion.
 Finally, we investigate the decoherence of an 
 accelerating qubit in Minkowski spacetime due to the Unruh effect. In
 this case decoherence is not due to fluctuations in the metric, but
 instead is caused by coupling (which we model with a standard
 Hamiltonian) between the qubit and the thermal cloud of Unruh
 particles bathing it. When the accelerating qubit is entangled with a
 stationary partner, the decoherence should induce a corresponding loss in
 teleportation fidelity.
\end{abstract}


\maketitle

\section{Introduction}

For much of its history, experimental observation of quantum
gravitational effects has been little more than an impossible dream
for general relativity. Yet recent theoretical developments have begun
to alter this view. New ideas in quantum
gravity, including string theory, are being explored, which, if
correct, would imply the presence of quantum gravitational behavior at length
scales much larger than Planckian~\cite{BHfactories,Jacobson,Yurtholo}.
Combined with emerging sensor and other technologies, these speculations
are raising the specter of near-term experimental tests for some of the
most fundamental manifestations of quantum spacetime structure.

An imperfect but illuminating analogy can be pursued in the physics
of fluids. Consider a hypothetical stage in the development of the
theory of fluids in which we are unaware of the atomic structure
of matter, while we understand the Navier-Stokes equations and are
somehow aware of Planck's constant, $\hbar$. In this analogy, the atomic
structure of fluids is the unknown holy grail of quantum fluid mechanics
which we are striving to discover with our limited knowledge. The
fundamental quantities characterizing a fluid classically can be taken
to be the speed of sound, $c_s$, and the density, $\rho$. Along with
$\hbar$, these quantities can be combined to define a length scale
\begin{equation}\label{lscale}
l_q = \left( \frac{\hbar}{\rho \, c_s} \right)^{\!\! \frac{1}{4}}
\end{equation}
which would seem to set the quantum regime where classical fluid
mechanics must break down. For typical fluids (e.g.\ water), $l_q$ is
of order $10^{-8} \, \mbox{cm}$;
it does a remarkably good job of predicting the atomic scale. But in
our hypothetical ignorance of atoms, we would be ill-advised to conclude
that quantum effects must be negligible all the way down to the length
scale $l_q$. Indeed, we know (in hindsight)
from kinetic theory that the true
length scale which sets the breakdown of classical fluid mechanics is
the correlation length, which is typically much larger than $l_q$.
Collective phenomena such as phase transitions may give rise to
correlation lengths that are macroscopically large. Even away from phase
transitions, correlation lengths are large enough to have effects
(such as Brownian motion) that are readily observable in experiments
performed far above the lengthscale $l_q$.

What this analogy teaches us is to be open to the possibility that, while
the Planck length sets the scale for quantum gravity, there may be
quantum effects of the unknown small-scale structure of spacetime
which are analogous to phase transitions or Brownian motion, and which
might be detectable at scales far above Planck. In this paper we will
explore one possible source for such effects:
quantum fluctuations in a background spacetime causing decoherence in
qubits kinematically coupled to that background. In a sense, such decoherence
would be analogous to Brownian motion caused by the small-scale structure
of a fluid. We will make very few
assumptions about the quantum theory of spacetime underlying the
fluctuations, apart from demanding that it results in a Hilbert space
structure for states of the metric which obeys the standard laws of
quantum mechanics. We will conclude the paper by discussing
a slightly different scenario for gravitational decoherence:
the decoherence of an accelerating qubit in Minkowski spacetime
due to the Unruh effect. In this case decoherence is not due to
fluctuations in the metric, but instead is caused by a coupling
(which we model with a standard Hamiltonian) between the qubit
and the thermal cloud of Unruh particles bathing it.

A separate set of motivations for studying gravitational decoherence
arises from relativistic quantum information theory,
a new field of research
that studies the properties of quantum information and
quantum communication as seen by observers
in moving frames. One of the most important ingredients of quantum
information theory is entanglement, and much of the interest so far
has been directed to its properties under Lorentz transformations. 
Czachor studied a version of the EPR experiment with relativistic
particles \cite{czachor97}, and Peres {\em et al.} demonstrated that
the spin of an electron is not covariant under Lorentz transformations 
\cite{peres02}. Furthermore, the effect of Lorentz transformations on
maximally spin-entangled Bell states in momentum eigenstates was
studied by Alsing and Milburn \cite{alsing02}, and Gingrich and Adami
derived the general transformation rules for the spin-momentum
entanglement of two qubits \cite{gingrich02}. Recently, these results
were extended to the Lorentz transformation of polarization entanglement
\cite{gingrich03}, and to situations where one observer is accelerated
\cite{alsing03}. Here, we take a look at a different aspect of
relativistic quantum information theory: the effect of
decoherence on a qubit due to quantum fluctuations in the metric. 

We will now turn our attention to the study of decoherence
in the four major paradigms of
relativity: First, we will calculate the effect of flat fluctuations
in the fabric of Minkowski spacetime on a linearly moving qubit. Next, we
put a qubit in orbit around a black hole. The mass fluctuations due to
the Hawking radiation induce fluctuations in the
Schwarzschild metric, which in turn couple to our qubit. Following
black holes, we
will study the interaction of a linearly moving qubit with coherent
and squeezed gravity waves. Coherent states correspond to classical
gravitational waves which are expected to arise from
astrophysical processes such as the inspiral of compact
binaries, while squeezing arises due to the expansion of the universe.
Finally, we study a related but qualitatively different decoherence phenomenon:
we derive the decoherence of a linearly accelerating qubit
in a bath of Unruh radiation. This treatment requires a
field-theoretical description of the thermal Unruh bath, while we
will be content with a first-quantized description
of the qubit, whose interaction with the quantum field of Unruh
particles we will model with a standard Hamiltonian.

First, we define a general quantum system to be in a state
$|\psi\rangle = \sum_k c_k |k\rangle$, where $\{ |k\rangle\}_k$ is a
complete set of (non-degenerate and discrete) eigenstates of the
Hamiltonian. The free evolution of $|\psi\rangle$ is given by 
\begin{equation}\label{first}
  |\psi(t)\rangle = \sum_k c_k e^{i E_k t / \hbar} |k\rangle \equiv
   \sum_k c_k e^{i\omega_k t} |k\rangle\; .
\end{equation}

We consider the situation where the quantum system moves along a geodesic
path in some fixed metric $g_0$, while the true metric
$g$ is subject to quantum fluctuations about $g_0$.
The proper time of the system is
denoted by $\tau$, and we have $|\psi(\tau)\rangle = \sum_k c_k
e^{i\omega_k\tau} |k\rangle$. The laboratory coordinates are $t$, $x$,
$y$, and $z$. We want to know the state of the quantum system
$|\psi(t)\rangle$ in laboratory coordinates. This depends on the
metric $g$:
\begin{equation}
  ds^2 = g_{\mu\nu}\, dx^{\mu} dx^{\nu} = d\tau^2\; ,
\end{equation}
where the last equality follows from the geodesic motion of the
quantum system.

Now suppose that the metric itself is a (highly delocalized) quantum
system in a state $\rho_g$. It is our central assumption that this
quantum metric behaves as a regular quantum system. In particular, we
assume that the superposition principle is valid for quantum states of
the metric. We initialize the state of our quantum system at time
$t=0$ in $|\psi(0)\rangle = \sum_k c_k |k\rangle$, and let it evolve
freely within the metric $\rho_G$ for a certain period. The state of
the system will generally become entangled with the metric, and we
wish to determine the reduced density matrix of the system at time
$t$.  

In general, the free evolution of an energy eigenstate $|k\rangle$ in a
particular quantum metric $|g_j\rangle$ is given by
\begin{equation}\label{interaction}
  |k\rangle |g_j\rangle \rightarrow e^{i\omega_k\tau_j} |k\rangle
   |g_j\rangle\; ,
\end{equation}
with $d\tau = dt\sqrt{g_{\mu\nu}\, \dot{x}^{\mu}\, \dot{x}^{\nu}}$, and
$\dot{x}^{\mu} \equiv dx^{\mu}/dt$. There is no back-action of the
state of the qubit onto the metric. Specifying the (fluctuating)
quantum metric and the geodesic motion of the quantum system allows us
to determine the reduced density matrix of the quantum system.

Rather than the general state in Eq.~(\ref{first}), we consider a qubit
in a state
\begin{equation}
  |\psi\rangle = \frac{1}{\sqrt{2}}\left( |0\rangle + e^{i\varphi}
   |1\rangle\right)\; .
\end{equation}
We could have chosen a more general state with differing relative
amplitudes, but it turns out that this state gives rise to all the
interesting physical features of the interaction model we study
here. Similarly, the general case of an $N$-level system does not give
rise to conceptually new physics.  

Next, we will consider the geodesic motion of a qubit in three different
fluctuating quantum spacetime metrics. We consider ``flat''
fluctuations in a two-dimensional Minkowski space in
Sec.~\ref{sec:flat}, mass fluctuations
in the Schwarzschild metric for circular orbits of the qubit in
Sec.~\ref{sec:bh}, and graviton-number fluctuations in coherent and
squeezed gravity waves in Sec.~\ref{sec:gw}.

\section{Fluctuations in flat spacetime}\label{sec:flat}

In this section we calculate the density matrix of a qubit that
experiences decoherence due to ``flat'' fluctuations in the Minkowski
spacetime metric. 

\subsection{The Minkowski metric}

First, consider a uniformly moving qubit with velocity
$dx/dt=v$ in the $x$-direction in a two-dimensional flat Minkowski
space. Let the state of the metric be given by
\begin{equation}\label{metric}
  \rho_g = \int d\vec{a}\, f(\vec{a})\,
  |g_{\vec{a}}\rangle\langle g_{\vec{a}}|\; ,
\end{equation}
with $\vec{a} = (a_1,a_2,a_3,a_4) \in \Bbb{R}^4$ and
$g_{\vec{a}}$ a particular metric given by  
\begin{equation}
  g_{\vec{a}} = 
  \begin{pmatrix}
   a_1 + 1 & a_2 \cr
   a_3 & a_4 + 1
  \end{pmatrix}
  \begin{pmatrix}
   -1 & 0 \cr
   0 & 1
  \end{pmatrix}
  \begin{pmatrix}
   a_1 + 1 & a_3 \cr
   a_2 & a_4 + 1
  \end{pmatrix}.
\end{equation}
This defines the ``flat'' fluctuations in the two-dimensional
Minkowski space. The distribution function $f(\vec{a})$ determines the
quantum fluctuations of the metric around $\vec{a}=0$. We assume that
the fluctuations are Gaussian:
\begin{equation}
  f(\vec{a}) = \exp\left( -\sum_{jk=1}^4 a_j A_{jk} a_k \right) \equiv \exp\left[ -(\vec{a}, A \vec{a}) \right]\; ,
\end{equation}
where $A$ is a positive symmetric variance matrix. The (dimensionless)
parameters specified by the matrix elements of $A$ would ideally be
derived from a complete quantum theory of gravity. In the absence of
such a theory, we would treat the $A_{ij}$ as arbitrary real parameters.

The proper time $\tau$ of the qubit in motion, given a particular
metric $g_{\vec{a}}$, is given by   
\begin{equation}\label{propertime}
  d\tau = dt \sqrt{g_{00} + 2 g_{0x}\, v + g_{xx}\, v^2}\; ,
\end{equation}
where
\begin{subequations}
\begin{eqnarray}
  g_{00} &=& a_1^2 - a_2^2 + 2 a_1 + 1 \\
  g_{0x} &=& a_1 a_3 - a_2 a_4 + a_3 - a_2 \\
  g_{xx} &=& a_3^2 - a_4^2 - 2 a_4 - 1 \; .
\end{eqnarray}
\end{subequations}
Furthermore, we assume linear motion, yielding $dx = v\, dt$, with
$\hbar = c = G = 1$. In the following, we will use $\Omega\equiv
\omega_1 - \omega_0$. 

\subsection{The decoherence model}

The procedure to calculate the decoherence due to the above
interaction of the qubit with the metric is as follows: first we write
the total state $\rho = |\psi\rangle\langle\psi|\otimes\rho_g$, and we
apply the interaction from Eq.~(\ref{interaction}). We then substitute
the expression from Eq.~(\ref{propertime}) into the resulting
state. In fact we will use the polynomial expansion of the square root
to second order. This is justified since the fluctuations are small.
Subsequently, we trace out the metric state, since we have no direct
access to its fluctuations. We can then evaluate the integral in
Eq.~(\ref{metric}), which yields an expression for the reduced density
matrix of the qubit.

The joint system of the qubit and the metric after the interaction of
Eq.~(\ref{interaction}) is in the state
\begin{eqnarray}
  \rho &=& \int_{{\Bbb{R}}^4} d\vec{a}\, e^{-(\vec{a},A\vec{a})}
  \left[ |0\rangle\langle 0| + e^{-i(\varphi + \Omega
  \tau)}|0\rangle\langle 
  1| \right.\cr && \qquad\qquad \left. + e^{i(\varphi + \Omega
  \tau)}|1\rangle\langle 0| + |1\rangle\langle 1| \right]\; . 
\end{eqnarray}
Substituting Eq.~(\ref{propertime}) into the above expression and
expanding to second order will yield a state $\alpha|0\rangle\langle 0| +
e^{-i\varphi}\beta^* |0\rangle\langle 1| + e^{i\varphi}\beta
|1\rangle\langle 0| + \alpha|1\rangle\langle 1|$, with
\begin{equation}
  \alpha = \int_{{\Bbb{R}}^4} d\vec{a}\, e^{-(\vec{a},A\vec{a})} =
  \frac{\pi^2}{\sqrt{\det A}}\; ,
\end{equation}
and 
\begin{equation}
  \beta = \int_{{\Bbb{R}}^4} d\vec{a}\, e^{-(\vec{a},A\vec{a}) +
  i\Omega\tau(\vec{a})}\; . 
\end{equation}
The proper time $\tau(\vec{a})$ is approximated by 
\begin{equation}\label{tau}
  \tau(\vec{a}) = t\left[ c_0 + (\vec{c}_1,\vec{a}) + (\vec{a},C_2
  \vec{a}) + O(\vec{a}^3) \right]\; , 
\end{equation}
with
\begin{subequations}
\begin{eqnarray}
  c_0(v) &=& 1 - \frac{v^2}{2} - \frac{v^4}{8} + \ldots = \sqrt{1-v^2} \equiv
  \gamma \\
  \vec{c}_1(v) &=& \left( 1-\frac{v^2}{2} \right) (1,-v,v,-v^2) \\
  C_2(v) &=& \frac{1}{2}
  \begin{pmatrix}
   \frac{v^2}{2} & v & \frac{v^3}{2} & v^2 \cr
   v   & 1 - \frac{3v^2}{2} & v^2 & -v-\frac{3v^3}{2} \cr 
   \frac{v^3}{2} & v^2 & \frac{v^4}{2} & v^3 \cr
   v^2 & -v-\frac{3v^3}{2} & v^3 & -v^2-\frac{3v^4}{2} \cr 
  \end{pmatrix} .
\end{eqnarray}
\end{subequations}

In order to calculate $\beta$, we collect all the second-order terms
$a_j a_k$ into the variance matrix $A$ (yielding a new positive
symmetric matrix $B = A + i\Omega t\, C_2$), and the first-order
terms are collected in a linear exponential:  
\begin{equation}\label{beta}
  \beta = e^{i\Omega t\gamma} \int_{{\Bbb{R}}^4}
  d\vec{a}\, e^{-(\vec{a},B\vec{a}) + (\vec{u},\vec{a})}\; ,
\end{equation}
where $\vec{u} = i\Omega t(1+v^2/2)(1,-v,v,-v^2)$. The overall
phase factor 
originates from the time dilation observed for a moving body with
velocity $v$. The integral in Eq.~(\ref{beta}) can be evaluated
formally to yield
\begin{equation}
  \int_{{\Bbb{R}}^4} d\vec{a}\, e^{-(\vec{a},B\vec{a}) +
  (\vec{u},\vec{a})} = \frac{\pi^2
  e^{(\vec{u},B^{-1}\vec{u})}}{\sqrt{\det B}}\; .
\end{equation}
For the integral to exist, the real part of the eigenvalues of $B$
must be strictly positive, which is ensured by $A$. 

Given a particular variance matrix $A$, we can calculate the
(normalized) state of the qubit as a function of the travelled
coordinate time $t$:
\begin{equation}\label{varphi}
  \rho(t) = \frac{1}{2}
  \begin{pmatrix}
   1 & \eta\, e^{i(\varphi'+\delta)}e^{-\Gamma^2} \cr 
   \eta\, e^{-i(\varphi'+\delta)}e^{-\Gamma^2} & 1
  \end{pmatrix}\; ,
\end{equation}
where $\varphi'=\varphi + \gamma\Omega t$ is the time-dilated
free evolution of the qubit, and $\eta\equiv \sqrt{\det A/\det
  B}$. Furthermore, we defined 
\begin{equation}
  \Gamma^2 \equiv \real[-(\vec{u},B^{-1}\vec{u})] \quad\text{and}\quad 
  \delta \equiv \imaginary[-(\vec{u},B^{-1}\vec{u})]\; .
\end{equation}

When the variance matrix is diagonal ($A = \sigma^{-2}\unity$), i.e.,
when there are no special correlations between the different
space-time dimensions, we can give a fairly straightforward expression
of the decohered qubit. Since the quantum fluctuations of the metric
are assumed small (Planck scale), the variance $\sigma$ is small, and
we can expand the solution around $\sigma=0$. Using Mathematica, we
find that 
\begin{equation}\label{Gamma}
  \Gamma = (1+v^2)\, \Omega t\, \sigma\; ,
\end{equation}
and 
\begin{equation}\label{delta}
  \delta = \frac{1}{4} v^2 (5 + 5 v^2 + 11 v^4 + 3 v^6)\, \Omega^3 t^3
  \sigma^4\; . 
\end{equation}

\begin{figure}[t]
  \begin{center}
        \epsfig{file=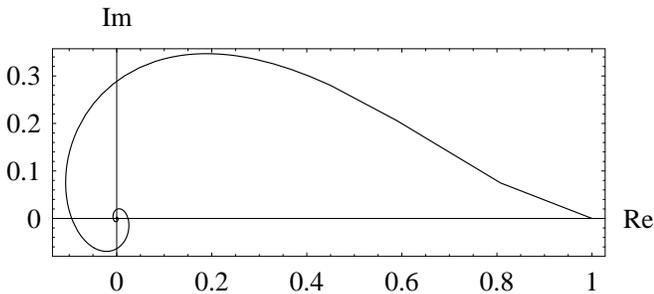, height=4cm}
  \end{center}
  \caption{The amplitude and phase of the off-diagonal element of the
    density matrix. The inward spiralling is due to the phase drift
    and the decoherence. The knee in the figure around the point
    $(\real,\imaginary)=(0.8,0.1)$ is due to the complex factor $\eta$.} 
  \label{fig:phase}
\end{figure}

Interestingly, there are three effects that alter the state of the
qubit. First, there is the expected decoherence $\exp(-\Gamma^2)$, the
argument of which scales quadratically with time $t$, frequency
difference $\Omega$, and variance $\sigma$. Secondly, we found a phase
drift $\delta$, due to the interaction with the quantum metric. This
effect scales with $(\Omega t)^3$ and $\sigma^4$, and it arises when
there  are off-diagonal terms in the matrix $B$, or, equivalently,
when there are $a_j a_k$ terms with $j\neq k$ in
Eq.~(\ref{tau}). Finally the factor $\eta$ behaves according to 
\begin{equation}
  \eta = \sqrt{\frac{\det A}{\det B}} \propto 1 + \frac{i}{2}
  (1+v^2)^2 \Omega t \sigma^2\; .
\end{equation}
The evolution of the complex off-diagonal element is shown in
Fig.~\ref{fig:phase}. 

The interaction of the quantum fluctuations of the metric with a qubit
generalizes directly to the case of an $N$-level system, or {\em
  quNit}. The density matrix will be an $N\times N$ matrix, and the
off-diagonal elements will be of the form $\eta\, e^{i(\varphi_{jk}' +
  \delta_{jk})} e^{-\Gamma_{jk}}$. We obtain $\varphi_{jk}'$,
$\delta_{jk}$, and $\Gamma_{jk}$ from Eqs.~(\ref{varphi}),
(\ref{Gamma}) and (\ref{delta}) by substituting $\Omega_{jk} \equiv
\omega_j - \omega_k$ for $\Omega$.

\subsection{Ensemble of qubits}

Note that the above calculation describes an
experiment in which the ensemble of qubits necessary
to statistically measure the reduced quantum state of the qubit is
obtained via repetition of each ensemble qubit's time
evolution through the same metric state Eq.~\ref{metric}.
What happens when we send an ensemble of $N$ qubits with uniform
velocity $v$, while their states are all subjected to the same fluctuations
in the Minkowski metric simultaneously?
Let $|s\rangle$ be an $N$-bit string, and
$0\leq n(s)\leq N$ the number of ones in $|s\rangle$. The evolution of
a bit string $|s\rangle$ on the metric $|g_{\vec{a}}\rangle$ in terms
of the proper time $\tau$ is given by  
\begin{equation}
  |s\rangle|g_{\vec{a}}\rangle ~\rightarrow~ \exp[i\, n(s)\, \Omega
   \tau_{\vec{a}}] |s\rangle|g_{\vec{a}}\rangle\; .
\end{equation} 
There are $2^N$ such strings, comprising a set ${\mathcal{S}}$,
and the density operator becomes a $2^N 
\times 2^N$ matrix: 
\begin{equation}
  \rho_N = \int_{{\Bbb R}^4} d\vec{a}\, e^{-(\vec{a},A\vec{a})} \left[
  \sum_{s,s'} e^{i[n(s)-n(s')\Omega t]} |s\rangle\langle s'| \right]\; .
\end{equation}

It is immediately obvious that the states most sensitive to this type
of decoherence is the maximally entangled GHZ state $|00\ldots 0\rangle
+ |11\ldots 1\rangle$, that is, states that have maximal $n(s)-n(s')$. 
As a direct consequence, these states can in principle be used to
detect the fluctuations \cite{lee02}.

Similarly, it is now also clear that such decoherence can be countered
by encoding quantum information in the decoherence-free subspace
spanned by the subsets of ${\mathcal{S}}$ for which
$n(s)-n(s')=0$. For a given $n(s)$, there are $\binom{N}{n(s)}$ such
states. This behaviour is reminiscent of the technique used in
Ref.~\cite{bartlett03}.

\section{Mass fluctuations in black holes}\label{sec:bh}

\subsection{The Schwarzschild metric}

Now, we will turn our attention to the situation of a qubit orbiting a
black hole, and we again assume that the metric is a quantum
mechanical object. The no-hair theorems for black-hole structure
suggest that quantum fluctuations in the metric will consist
of fluctuations in the mass of the black hole (hence we neglect the
fluctuations in angular momentum, and assume that charge fluctuations
are forbidden in quantum gravity because of super-selection rules).

For simplicity, we assume
that the orbit of the qubit is circular. The Schwarzschild metric of
a given mass $M$ is given by  
\begin{equation}
  ds^2 = g_M \equiv -\left( 1-\frac{2M}{r} \right) dt^2 +
  \frac{dr^2}{1-\frac{2M}{r}} + r^2 d\Omega^2\; ,
\end{equation}
where $d\Omega^2 \equiv d\theta^2 + \sin^2\theta\, d\phi^2$.

The evolution of energy eigenstates is again given by
Eq.~(\ref{interaction}). We need to express $\tau$ in terms 
of $t$. To this end, we start from the Hamiltonian for geodesic motion
in the Schwarzschild background
\begin{eqnarray}\label{hamilton}
  H &=& \frac{1}{2} g^{\mu\nu} p_{\mu} p_{\nu} \cr
    &=& -\frac{p_0^2}{2-\frac{4M}{r}} + \left(
    1-\frac{2M}{r} \right) \frac{p_r^2}{2} + \frac{p_{\theta}^2}{2r^2} +
    \frac{p_{\phi}^2}{2r^2\sin^2\theta} \cr
    &=& -\frac{E^2}{2\left( 1-\frac{2M}{r}\right)} + \frac{1}{2}
    \left( 1-\frac{2M}{r} \right)p_r^2 + \frac{L^2}{2r^2} \cr 
    &=& -\frac{1}{2}\; .
\end{eqnarray}
Here we have assumed without loss of generality
that the orbital plane has $\theta=\pi/2$ and
thus by spherical symmetry $p_{\theta}=0$. Furthermore,
we introduced the energy $E=-p_0$ and the angular momentum
$L=p_{\phi}$. The last equality holds since the four-velocity $v
= dx/d\tau$ is normalized: $v^{\mu} v_{\mu} = g^{\mu\nu} p_{\mu}
p_{\nu} = -1$. Furthermore, we have
\begin{equation}
  p_0 = g_{00} v^0 = -\left( 1-\frac{2M}{r} \right) \frac{dt}{d\tau} =
  -E\; ,
\end{equation}
which gives us an expression for $\tau$ in terms of $t$ and $E$. We now need
to determine $E$.

>From Eq.~(\ref{hamilton}) and $p_r=0$ (since we consider a circular orbit),
we derive
\begin{equation}
  E^2 = \left( 1-\frac{2M}{r} \right) \left( 1+\frac{L^2}{r^2} \right)
  \equiv V(r) \; .
\end{equation}
For stable orbits we need the potential $V(r)$ to be a minimum:
\begin{equation}\label{conditions}
  \left.\frac{\partial V}{\partial r}\right|_{r_0} = 0 
  \quad\text{and}\quad
  \left.\frac{\partial^2 V}{\partial r^2}\right|_{r_0} > 0 \; .
\end{equation}
This yields
\begin{equation}
  \left( 1-\frac{2M}{r_0} \right) \frac{L^2}{r^3_0} - \frac{M}{r^2_0}
  \left( 1+\frac{L^2}{r^2_0} \right) = 0\; ,
\end{equation}
or
\begin{equation}
  L^2 = \frac{Mr_0^2}{r_0 - 3M}\; 
\end{equation}
which makes manifest the well-known fact
that there are no stable circular orbits inside the critical
radius $r = 3 M$. For any $r_0 > 3M$, a circular orbit
at $r=r_0$ can be found with
the above value of $L$. On such an orbit, the energy $E$ is determined
by substituting $L^2$ into $V(r_0)$; from Eq.\,(27) we find that
\begin{equation}
  d\tau^2 = \left( 1-\frac{3M}{r_0} \right) dt^2\; .
\end{equation}

We will assume that the black hole is in thermal equilibrium with the
thermal Hawking radiation it emits at the temperature $T_H$. This is
equivalent to assuming that the black hole is inside a conducting sphere
with perfectly reflecting walls placed at a large radius away from the
horizon. The quantum state of the metric can then be written as in
Eq.~(\ref{metric}) in the form
\begin{equation}
  \rho_g = \int dM \, \exp \left[ -\frac{(M-M_0 )^2}{2 \sigma_M^2}
\right]
  |g_M \rangle\langle g_M |\; ,
\end{equation}
where $M_0$ denotes the classical mass of the hole, about which the
mass $M$ fluctuates in equilibrium with the thermal bath at temperature
$T_H$. The fluctuations are Gaussian (in accordance with the
thermodynamic limit) around $M_0$. More precisely, a general canonical
ensemble with Hamiltonian $H$ has energy fluctuations given by
\begin{equation}
\langle H^2 \rangle - \langle H \rangle^2
= \frac{\partial^2}{\partial \beta^2} \log Z =
-\frac{\partial}{\partial \beta} \langle H \rangle \; ,
\end{equation}
where $Z = \int e^{- \beta H} $ is the partition function,
$\beta=1/(k_B T)$ is the inverse temperature, and
$\langle \cdots \rangle$ denotes the canonical ensemble average;
thus, e.g., $\langle H \rangle $ is equal to the internal energy $U$.
According to our assumption as explained below,
$\sigma^2_M$ will be given by
\begin{equation}
\sigma_M^2 = \langle H^2 \rangle_{T_H} - \langle H {\rangle^2}_{T_H} \; ,
\end{equation}
i.e.\ equal to the energy fluctuations
of a Bose gas at thermal equilibrium with black-body radiation at the
Hawking temperature.

In order to find the effect of thermal fluctuations in the black-hole
mass on the evolution of coherence in the state of our qubit in circular
orbit at $r=r_0$, we
calculate the off-diagonal element of the qubit's density matrix:
\begin{eqnarray}
  \beta &=& \int_{-\infty}^{+\infty} dM \, \exp\left[
  -\frac{(M-M_0)^2}{2\sigma^2_M} \right]  e^{i \Omega
  t\sqrt{1-\frac{3M}{r_0}}} \cr
  &=& \sqrt{\frac{9M_0^2 \Omega t}{8r_0^2 + 18 i(\Omega)^2
  t \sigma_M^2}}\; \exp\left(-\Gamma^2 + i\delta\right)\; ,
\end{eqnarray}
where 
\begin{equation}\label{bhgamma}
  \Gamma = \frac{9\sqrt{2}}{8} \left(\frac{3M_0 + 2r_0}{r_0^2}\right)
  \Omega t\, \sigma_M\; ,
\end{equation}
and
\begin{eqnarray}\label{bhdelta}
  \delta &=& \Omega t \sqrt{1-\frac{3M_0}{r_0}} \cr && \qquad +
  \frac{81}{128}\left(\frac{3M_0 + 2r_0}{r_0^6}\right)^2
  \Omega^3 t^3\, \sigma_M^4\; .
\end{eqnarray}
The first term of the phase drift is the regular
relativistic effect, and the second term is the
variational phase drift due to the quantum fluctuations in the mass of
the black hole. The off-diagonal element of the density matrix evolves
similar to the one in the previous section (see Fig.~\ref{fig:phase}).

\subsection{Mass fluctuations}

Now we seek an expression for $\sigma_M$. We know that the Hawking
temperature of a black hole is proportional to the surface gravity
$\kappa$ \cite{wald94}:
\begin{equation}
  kT_H = \frac{\hbar}{c}
\frac{\kappa}{2\pi} = \frac{\hbar c^3}{8\pi G M}\; ,
\end{equation}
where we have used the fact
that a non-rotating neutral black hole of mass $M$ has
$\kappa=(4GM/c^4)^{-1}$, and therefore ${\beta}^{-1}=
kT_H = \hbar c^3 / (8\pi G M )$
\cite{wald84}. For clarity we will keep the physical
constants $\hbar$, $c$ and
$G$ in arbitrary units throughout this subsection.

The Hawking radiation has a perfect black-body spectrum.
We will assume that the fluctuations in the mass of the hole are
identical to the fluctuations in the average energy $U$ of a black body of
surface area $4 \pi G^2 {M_0}^2 / c^4$ (the area of the black-hole horizon)
radiating at the Hawking temperature $T_H$. The average
internal energy of a
black body (Boson gas) of volume $V$ at temperature $T$ is given by
\begin{equation}
\langle H \rangle =
U = \frac{4 \sigma}{c}
\, V \, T^4 = \frac{\pi^2 {k_B}^4}{15 \hbar^3 c^3} V \, T^4
= \frac{\pi^2 }{15 \hbar^3 c^3 } \frac{V}{\beta^4} \; ,
\end{equation}
where $\sigma$ is the Stefan-Boltzmann constant.
According to Eqs.\,(34)--(35)
\begin{equation}
c^4 \; \sigma_M^2 = \langle H^2 \rangle - \langle H \rangle^2
=  \frac{4 \pi^2 }{15\hbar^3 c^3} \frac{V}{\beta^5} \; .
\end{equation}
Note that the black-body energy $U$ itself is {\em not}
equal to the classical hole mass $M_0$; it is
only the mean-square fluctuations $\sigma_M^2$ about $M_0$ that we
assume are identical to the thermal fluctuations in $U$ at the Hawking
temperature.
Substituting the Hawking value $\beta = 8 \pi G M_0/(\hbar c^3)$,
and the geometric identity
$V=(32 \pi /3) G^3 {M_0}^3/c^6$ in Eq.\,(41), we finally obtain
\begin{equation}
\sigma_M^2 = \frac{\hbar^2 c^2}{11520 \, \pi^2 G^2} \frac{1}{{M_0}^2}
=\frac{{m_p}^2}{11520 \, \pi^2} \left( \frac{m_p}{M_0} \right)^2 \; ,
\end{equation}
where $m_p = \sqrt{\hbar c /G}$ is the Planck mass. Going back to
geometric (Planck) units (where Planck mass is unity),
we can write Eq.\,(42) in the form
\begin{equation}
\sigma_M^2 =
\frac{1}{11520 \, \pi^2}
\left( \frac{1}{M_0} \right)^2 \; ,
\end{equation}
which is the value that needs to be substituted in Eqs.\,(33)--(38). 

To give a feeling for the order of magnitude of this decoherence
effect, we note that in order to have the off-diagonal elements of the qubit state
decohere to (an absolute value of) $1/e$, the value of $\Gamma$
[cf.\ Eq.\,(37)] must be 1. Given a black hole
of one solar mass ($10^{38}$ Planck masses), and a qubit
with transition frequency
$\Omega=10^{15}$Hz in the lowest stable orbit ($r_0 = 3 M_0$), the
(asymptotic, Minkowskian) coordinate
time necessary to reach this level of decoherence
is $10^{13}$ years. To get a decoherence effect
of the same magnitude over a
time interval of one year, the black hole must be no more
massive than 10 kilograms. On the other hand, if we take the suggestions of
microscopic black-hole production in the hadron colliders currently
under construction seriously~\cite{BHfactories},
their decoherence effect on nearby quantum
states should be considerably larger.

\section{Gravitational Waves}\label{sec:gw}

In this section, we will consider the interaction of a qubit with
coherent and squeezed graviton fields through the proper time
evolution of the qubit. In the``transverse-traceless" gauge, the metric for
classical gravitational waves travelling in the $z$-direction is given
by \cite{yurtsever88} 
\begin{eqnarray}
  ds^2 &=& -du\, dw + (1+h_+^c)^2 dx^2 + (1-h_+^c)^2 dy^2 \cr && +
  2 h_{\times}^c dx\, dy\; , 
\end{eqnarray}
where $u=t-z$ and $w=t+z$, and $h^c_{+,\times}$ the classical
amplitudes of the gravitational wave in the $+$ and the $\times$
polarization. Furthermore, the qubit moves in the $xy$ plane with
velocity $v$ such that  
\begin{equation}
  \frac{du}{dt} = \frac{dw}{dt} = 1\; , \quad \frac{dx}{dt} =
  v\sin\theta\; , \quad \frac{dy}{dt} = v\cos\theta\; .
\end{equation}
This defines $\theta$. Expanding the metric to linear order in
$h_+^c$, the proper time of the qubit is then given by ($d\tau^2 = ds^2$):
\begin{equation}
  \tau = t \sqrt{(1-v^2)- 2 h_+^c v^2 \cos 2\theta + h_{\times}^c
  v^2 \sin 2\theta}\, .
\end{equation}
When we quantize the (weak) gravitational wave, the metric can be
approximated by 
\begin{eqnarray}
  ds^2 &=& -du\, dw + (1+2n\,h_+) dx^2 + (1-2n\,h_+) dy^2 \cr && 
  + 2m\,h_{\times} dx\, dy\; , 
\end{eqnarray}
where $n,m$ are the graviton numbers for $h_+$ and $h_{\times}$
respectively. The classical metric is then retrieved by setting 
\begin{equation}
  h_+^c \equiv \langle n\rangle h_+ \quad\text{and}\quad h_{\times}^c
  \equiv \langle m\rangle h_{\times}\; .
\end{equation}
Here $\langle n\rangle$ and $\langle m\rangle$ are the average
graviton numbers. The $h_+$ and $h_{\times}$ denote a distribution
functions over a set of frequencies, and may be a function of $u$ and
$w$. 

The proper time of the qubit interacting with $n$ $+$-polarized and
$m$ $\times$-polarized gravitons, expressed in the standard coordinates,
then becomes 
\begin{equation}
  d\tau_{nm} = dt \sqrt{(1-v^2)- 2n\, h_+ v^2 \cos
  2\theta + m\, h_{\times} v^2 \sin
  2\theta} \, .
\end{equation}
Suppose that we deal only with plane waves of specific frequencies
$\omega_+$ and $\omega_{\times}$ such that $h_+(t,z) = h_+
\cos\omega_+ t$ and $h_{\times}(t,z) = h_{\times} \cos(\omega_{\times}
t + \varphi)$. We can then approximate this expression using
$\sqrt{a+b} \approx \sqrt{a} + b/(2\sqrt{a})$:
\begin{eqnarray}
  \tau_{nm} &=& \gamma t - n \int_0^t dt'\, 2\, h_+ v^2 \cos
  2\theta \cos\omega_+ t' \cr && \quad + m \int_0^t dt'\, h_{\times} v^2 \sin
  2\theta  \cos (\omega_{\times} t' + \varphi)\, .
\end{eqnarray}
with $\gamma \equiv \sqrt{1-v^2}$. We can compactify this expression
and write $\tau_{nm} = \gamma t - n \tau_+ + m \tau_{\times}$, where 
\begin{subequations}\label{grtau}
\begin{eqnarray}
  \tau_+ &=& \frac{2 h_+ v^2 \cos 2\theta \sin\omega_+
  t}{\gamma\omega_+}\; , \\
  \tau_{\times} &=& \frac{h_{\times} v^2 \sin 2\theta
  \sin(\omega_{\times} t + \varphi)}{\gamma\omega_{\times}}\; . 
\end{eqnarray}
\end{subequations}

Let's consider quantized gravitational waves in the two polarization
modes $h_+$ and $h_{\times}$. That is, we can write the state of the
mode in terms of bosonic creation and annihilation operators
$\hat{h}_j^{\dagger}$ and $\hat{h}_j$, where $j\in\{+,\times\}$, and  
\begin{equation}
  [\hat{h}_j,\hat{h}_k^{\dagger}] = \delta_{jk} \quad\text{and}\quad
  [\hat{h}_j,\hat{h}_k] = [\hat{h}_j^{\dagger},\hat{h}_k^{\dagger}] =
  0\; .
\end{equation}
We will now use this formalism to study the effect of different
quantum states of gravitational radiation on the qubit.

\subsection{Coherent gravitational waves}

The most classical states of a bosonic field are the coherent states,
so it is natural to study coherent gravitational waves
\cite{lovas99}. We are looking 
at single modes of a single frequency. Analogous to optical coherent
states we define the coherent state of the (polarized) gravitational
wave with amplitudes $\eta_j$ ($j\in\{+,\times\}$) as
\begin{equation}
  |\eta_j\rangle = e^{-\eta_j^2/2} \sum_{n=0}^{\infty} \frac{\eta_j^n
   \hat{h}_j^{\dagger n}}{n!} |0\rangle = e^{-\eta_j^2/2}
   \sum_{n=0}^{\infty} \frac{\eta_j^n}{\sqrt{n!}} |n\rangle \; ,
\end{equation}
where $|n\rangle$ is the $n$-graviton Fock state obtained by 
\begin{equation}
  \hat{h}_j^{\dagger} |n\rangle_j = \sqrt{n+1} |n+1\rangle_j
  \quad\text{and}\quad
  \hat{h}_j |n\rangle_j = \sqrt{n} |n-1\rangle_j\; .
\end{equation}
Here, we have chosen $\eta_+, \eta_{\times}$ real without loss of
generality. 

When a qubit interacts with the gravitational wave, we write the state
of the joint system as
\begin{equation}\label{coherentstate}
  \frac{|0\rangle + |1\rangle}{\sqrt{2}} |\eta_+,\eta_{\times}
  \rangle \rightarrow \sum_{n,m=0}^{\infty} \frac{|0\rangle +
  e^{i\Omega\tau_{nm}}|1\rangle}{\sqrt{2}} \frac{\eta_+^n
  \eta_{\times}^m}{\sqrt{n! m!}} |n,m\rangle\; .
\end{equation}
Tracing out the gravitational wave then leaves us with a qubit in the
state 
\begin{equation}\label{cohout}
  \rho = \frac{1}{2}
  \begin{pmatrix}
   1 & \beta \cr
   \beta^* & 1
  \end{pmatrix}\; ,
\end{equation}
where 
\begin{equation}
  \beta \propto \sum_{n,m=0}^{\infty}
  \frac{\eta_+^{2n} e^{-in\Omega \tau_+}}{n!}~ \frac{\eta_\times^{2m}
  e^{-im\Omega \tau_{\times}}}{m!} 
\end{equation}
up to a constant $\exp[i\gamma\,\Omega t - \eta_+^2 -
\eta_{\times}^2]$. This  yields
\begin{equation}
  \beta = e^{i\gamma\,\Omega t} \exp\left[\eta_+^2 (e^{-i\Omega\tau_+}-1)
  + \eta_{\times}^2 (e^{-i\Omega\tau_{\times}}-1)\right].
\end{equation}
Expanding the the phase factor according to $e^{ix} = \cos x + i \sin
x$, we can formally rewrite $\beta = e^{-\Gamma + i \delta}$, with 
\begin{eqnarray}
  \Gamma &=& \eta_+^2\left( 1 - \cos\phi_+ \right) +
  \eta_{\times}^2\left( 1 - \cos\phi_{\times}  \right) \cr
  &\approx& \frac{1}{2}\left( \eta_+^2\phi_+^2 +
  \eta_{\times}^2\phi_{\times}^2 \right)  \; ,
\end{eqnarray}
and 
\begin{eqnarray}
  \delta &=& \Omega t \gamma - \eta_+^2 \sin\phi_+ + \eta_{\times}^2
  \sin\phi_{\times} \cr
  &\approx& \Omega t \gamma - \eta_+^2 \phi_+ + \eta_{\times}^2
  \phi_{\times}\; .  
\end{eqnarray}
Here, we defined
\begin{subequations}
\begin{eqnarray}
  \phi_+ &=& \frac{\Omega}{\omega_+}\; \frac{2 h_+ v^2 \cos 2\theta
  \sin\omega_+ t}{\gamma} \quad\text{and}\quad\\
  \phi_{\times} &=& \frac{\Omega}{\omega_{\times}}\; \frac{h_{\times}
  v^2 \sin 2\theta \sin(\omega_{\times} t + \varphi)}{\gamma}\; .  
\end{eqnarray}
\end{subequations}
This means that there is a {\em periodic} decoherence of the qubit
under influence of the gravitational wave. Indeed, the restoration of
coherence is induced mathematically by the occurrence of a phase
factor in the argument of the exponential. Similarly, the induced
phase drift $\delta$ is periodic in time. However, since $\phi_+$ and
$\phi_{\times}$ are themselves periodic functions of the laboratory
time $t$ with a small amplitude [see Eq.~(\ref{grtau})], the periodicity
of the decoherence and the phase drift will be extremely difficult to
observe in practice, due to their extremely small amplitude. 

However, in practice the qubit will not be interacting with a plane
wave, but with a pulse of finite extension. If we consider a simple
coherent Gaussian wave packet, the gravitational wave in one
polarization can be written as 
\begin{equation}
  |\Phi(t)\rangle = \frac{e^{-\eta^2/2}}{\sqrt[4]{\pi\sigma^2}}
   \sum_{n=0}^{\infty} \int_0^{\infty} d\omega\, e^{-
   \frac{(\omega-\omega_0)^2}{2\sigma^2} + i\omega t}
   \frac{\eta^n}{\sqrt{n!}} |n_{\omega}\rangle\; . 
\end{equation}
For simplicity we have included the dependence of $\eta$ on $\omega$
in the Gaussian function.

\begin{figure}[b]
  \begin{center}
        \epsfig{file=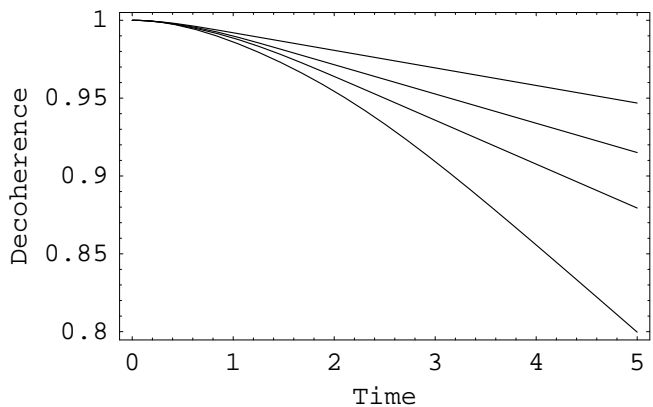,height=5.5cm}
  \end{center}
  \caption{The behaviour of ${\rm Tr}(\rho_{\rm out}^2)$ for different
    $\sigma$. When $\sigma$ becomes larger, the trace remains closer
    to one. The values of $\sigma$ are 1, 2, 3, and 5.} 
  \label{fig:coherent}
\end{figure}

Under the evolution $|k\rangle \otimes |n\rangle_g \rightarrow
e^{i\Omega_k\tau_n} |k\rangle \otimes |n\rangle_g$ the metric and the
qubit state $|\psi\rangle = (|0\rangle + |1\rangle)/\sqrt{2}$ become
entangled. When we trace out the gravitational wave, the state of the
qubit then becomes
\begin{widetext}
\begin{eqnarray}
  \rho_{\rm out} &=& {\rm Tr}_g \Biggl\{
  \frac{e^{-\eta^2}}{\sqrt{4\pi\sigma^2}} 
  \sum_{n,m=0}^{\infty} \int_{0}^{\infty} d\omega\, d\omega'\,
  \exp\left[ -\frac{(\omega-\omega_0)^2}{2\sigma^2}
  -\frac{(\omega'-\omega_0)^2}{2\sigma^2} - i(\omega - \omega')t
  \right] \frac{\eta^{n+m}}{\sqrt{n!\, m!}} |n_{\omega}\rangle \langle
  m_{\omega'}| \Biggr. \cr
  && \qquad\quad \Biggl. \otimes \left(
  |0\rangle\langle 0| + e^{-i\Omega\tau_n(\omega)} |0\rangle\langle 1|
  + e^{i\Omega\tau_n(\omega)} |1\rangle\langle 0| + |1\rangle\langle
  1| \right) \Biggr\} \cr
  &=& \frac{e^{-\eta^2}}{\sqrt{4\pi\sigma^2}} \sum_{n=0}^{\infty}
  \int_0^{\infty} d\omega\, e^{-\frac{(\omega-\omega_0)^2}{\sigma^2}}
  \frac{\eta^{2n}}{n!} \left( |0\rangle\langle 0| +
  e^{-i\Omega\tau_n(\omega)} |0\rangle\langle 1| +
  e^{i\Omega\tau_n(\omega)} |1\rangle\langle 0| + |1\rangle\langle 1|
  \right) =
  \begin{pmatrix}
   \alpha & \beta^* \\
   \beta & \alpha 
  \end{pmatrix}\; , 
\end{eqnarray}
\end{widetext}
where $\tau_n (\omega) = n \tau_k (\omega)$ with $\tau_k$ given by
Eq.~(\ref{grtau}) and $\omega = \omega_k$. We can calculate the matrix
elements of $\rho_{\rm out}$:
\begin{equation}
  \alpha = \frac{e^{-\eta^2}}{\sqrt{4\pi\sigma^2}} \sum_{n=0}^{\infty}
  \int_0^{\infty} d\omega\, \frac{\eta^{2n}}{n!} e^{-\frac{(\omega -
  \omega_0)^2}{\sigma^2}} = \frac{1}{2}\; ,
\end{equation}
and
\begin{equation}
  \beta = \frac{e^{-\eta^2}}{\sqrt{4\pi\sigma^2}} \sum_{n=0}^{\infty}
  \int_0^{\infty} d\omega\, \frac{\eta^{2n}}{n!} e^{-\frac{(\omega -
  \omega_0)^2}{\sigma^2} + in\mu\frac{\Omega}{\omega} \sin \omega t}\; ,
\end{equation}
with $\mu = 2 h v^2 \cos 2\theta / \gamma\nu$. In order to calculate
$\beta$, we note that $n\mu$ is of the order of the classical
amplitude of the gravitational wave, i.e., $10^{-17}$ at best. We can
therefore approximate the imaginary part of the exponential to the
first few orders, using $e^x \simeq 1+x+x^2/2$ ($x\ll 1$). We need the
second order in the approximation of the exponential, because it is
going to contribute to the decoherence [${\rm Tr}(\rho_{\rm out}^2)$]
to the same order as the first term: 
\begin{widetext}
\begin{equation}
  \beta \simeq \frac{1}{2} + \frac{i \eta^2 \mu}{\sqrt{4\pi\sigma^2}}
  \int_0^{\infty} d\omega\, \frac{\Omega}{\omega} \sin\omega t\;
  e^{-\frac{(\omega - \omega_0)^2}{\sigma^2}} 
   - \frac{\eta^4 \mu^2}{2\sqrt{4\pi\sigma^2}}
  \int_0^{\infty} d\omega\, \left(\frac{\Omega}{\omega} \sin\omega t\right)^2
  e^{-\frac{(\omega - \omega_0)^2}{\sigma^2}}.
\end{equation}
The expression $\eta^2 \mu$ is on the order of the classical
gravitational wave amplitude $h^c$, with $\eta\gg 1$ and $\mu\lll
1$. Now, $\beta$ can be written as 
\begin{equation}
  \beta = \frac{1}{2} \left\{ 1 + \frac{i h^c\Omega
  \sqrt{\pi}}{2\sigma}\; {\rm erf} \left( \frac{\sigma t}{2} 
  \right) -
  \frac{h^{c\,2}\Omega^2\sqrt{\pi}}{2\sigma} \left[
  \frac{e^{-t^2\sigma^2}- 1}{\sqrt{\pi}\sigma} + t\; {\rm erf}\left(
  \sigma t \right)\right]   \right\}\; .   
\end{equation}
The imaginary part of $\beta$ is shown in Fig.~\ref{fig:coherent}. The
magnitude of the decoherence [given by ${\rm Tr}(\rho_{\rm out}^2)$]
can now be calculated:
\begin{equation}
  {\rm Tr}(\rho_{\rm out}^2) = 1 -
  \frac{h^{c\,2}\Omega^2\pi}{2\sigma^2} \left[ \frac{e^{-\sigma^2 t^2}-
  1}{\pi} + \frac{\sigma\, t}{\sqrt{\pi}}\, {\rm erf}(\sigma\, t) -
  \frac{1}{2} {\rm erf} \left( \frac{\sigma\, t}{2} \right)^2
  \right]\; .
\end{equation}
The general behaviour of the decoherence is shown in
Fig.~\ref{fig:coherent}. 
\medskip
\end{widetext}

\subsection{Squeezed gravitational waves}

Now let's look at squeezed gravity waves. These states are expected to
arise due the expansion of the universe \cite{grishchuk90}. Suppose
the squeezed gravitational wave in one polarization mode (for
simplicity) has the following graviton-number expansion: 
\begin{equation}
  |\zeta\rangle_g = e^{\frac{\zeta}{2} \hat{h}^{\dagger 2} -
   \frac{\zeta^*}{2} \hat{h}^2} |0\rangle = \sum_{n=0}^{\infty}
   \frac{\tanh^n |\zeta| \sqrt{(2n)!}}{2^n\, n! \cosh|\zeta|} |2n\rangle\; ,
\end{equation}
where $\zeta$ is the complex squeezing parameter, which depends on the
rate of expansion of the universe. 

A squeezed gravitational wave packet can be defined as ($\tanh|\zeta|
\equiv r$):
\begin{equation}
  |\zeta(t)\rangle =
   \frac{\sqrt[4]{1-r^2}}{\sqrt[4]{\pi\sigma^2}} 
   \sum_{n=0}^{\infty} \int_0^{\infty} d\omega\, \frac{r^n
   \sqrt{(2n)!}}{2^n n!} e^{- \frac{(\omega - \omega_0)^2}{2
   \sigma^2}} |2n_{\omega}\rangle\; . 
\end{equation}
Using the techniques of the previous section, we can express the state
of the qubit again in the form of  
\begin{equation}
  \rho = \frac{1}{2} 
  \begin{pmatrix}
    1 & \beta \cr
    \beta^* & 1
  \end{pmatrix}\; ,
\end{equation}
where 
\begin{eqnarray}
  \beta &=& \sqrt{\frac{1-r^2 }{4\pi\sigma^2}} \int_0^{\infty} \frac{e^{-
  \frac{(\omega - \omega_0)^2}{\sigma^2}}
  d\omega}{\sqrt{1-e^{i\phi(\omega)} r^2}} \cr 
  &=& \frac{1}{2} \left\{ 1 + i\mu\Omega \frac{\sqrt{\pi}
  r^2}{2\sigma^2 (1-r^2)}\; {\rm erf} \left( \frac{\sigma t}{2}
  \right) - \mu^2\Omega^2 \right. \cr && \left. \times  \frac{r^2 (2+r^2)}{8
  (1-r^2)^2 \sigma^2}\left[ e^{-\sigma^2 t^2}-1+\sqrt{\pi}\sigma t\,
  {\rm erf}\left(\frac{\sigma t}{2} \right) \right] \right\} , \cr && 
\end{eqnarray}
where we defined $\phi(\omega) = \mu\Omega \sin\omega t/\omega$. The
off-diagonal elements of the density matrix therefore exhibit the
same behaviour as in the coherent case. The field amplitude of the
squeezed gravitational waves are now proportional to $\mu
r^2/(1-r^2)$, where $0\leq r\leq 1$ is the squeezing parameter. The
decoherence with time is again given in Fig.~\ref{fig:coherent}.

\section{Decoherence via Unruh radiation}\label{sec:unruh}

This section is different from the others in that we will not
consider a fluctuating quantum metric, but rather an accelerated qubit
in Minkowski spacetime interacting with a
thermal bath of Unruh particles. The quantum field whose states contain
the Unruh particles and which interacts with the qubit
will be assumed to be a real scalar field for
simplicity, but qualitatively similar results are valid for any linear
field independently of spin or charge. When the qubit is an atomic system,
the relevant quantum field is that of photons in flat spacetime.

\subsection{Acceleration in flat space-time}

\begin{figure}[t]
  \begin{center}
        \epsfig{file=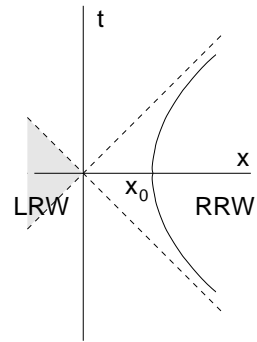, height=4.5cm}
  \end{center}
  \caption{The world line of an accelerated observer. LRW and RRW
    denote the left- and right-Rindler wedges, respectively. They are
    the spacelike separated regions with respect to the origin. The
    accelerated observer intersects the $x$-axis at $x_0$.} 
  \label{fig:unruh}
\end{figure}

We closely
follow the treatment given in~\cite{UnruhWald} of the behavior of an
accelerating particle detector in Minkowski spacetime. Accordingly, we
consider a qubit in uniform acceleration along the worldline (see
Fig.~\ref{fig:unruh}): 
\begin{equation}\label{rindler}
 t = \frac{1}{a} \sinh (a \tau ) \quad\text{and}\quad
 x = \frac{1}{a} \cosh (a \tau ) \; ,
\end{equation}
where $\tau$ denotes proper time along the worldline, and
$a$ is the magnitude of the acceleration. As in all standard treatments
of the Unruh effect, we envision a congruence of uniformly accelerating
world lines, where the worldline crossing the $x$-axis at
$x=x_0$ has acceleration $1/ x_0$. The coordinate origin
can be adjusted so that our given qubit's worldline crosses the $x$-axis at
$x=1/a$ in accordance with Eq.\,(\ref{rindler}). We will assume that the qubit
interacts with a real, massless scalar field $\phi$ propagating
on Minkowski spacetime in just the same way as a ``particle detector"
for $\phi$ would. Accordingly, we will assume a model interaction
Hamiltonian 
\begin{equation}\label{gwham}
 H_I (t) = \epsilon (t) \int_{\Sigma}
 \hat\phi (\vec{x}, t) \left[ \psi (\vec{x}) \hat{b} + {\psi}^\ast (\vec{x}) 
 \hat{b}^{\dagger} \right] \sqrt{-g} \, d^3 x \; .
\end{equation}
Here the integration is over the global spacelike
Cauchy surface $\Sigma = \{ t = \mbox{constant} \}$ in Minkowski
spacetime, and $\epsilon (t)$ is a coupling constant which is explicitly time
dependent to allow the acceleration to vanish outside a finite
time interval; we will assume that $\epsilon (t)$
is a constant, $\epsilon (t) \equiv \epsilon$, within a finite
interval in $t$ of duration $\Delta$ and zero outside that interval.
The function $\psi (\vec{x})$ is part of the coupling constant. It is a smooth
function that vanishes outside a small volume around the qubit, and
that models the finite spatial range of the interaction.

The field operator $\hat\phi (\vec{x}, t)$ can be expanded
either as the mode sum
\begin{equation}\label{grfield}
 \hat\phi (\vec{x} ,t) = \sum_i \left[ u_i (\vec{x} ,t) \,
 \hat{a}_M( u_i ) + {u_i (\vec{x}, t)}^{\ast} \hat{a}_M^{\dagger}
 (u_i) \right] \; ,
\end{equation}
where $\{ u_i \}$ is an orthonormal basis of positive-frequency
(with respect to the timelike Killing field $\partial / \partial t$)
solutions of the (Klein-Gordon) field equation for $\phi$
on Minkowski spacetime
and $\hat{a}_M ^{\dagger} (u_i )$ and $\hat{a}_M (u_i )$ denote the 
corresponding
creation and annihilation operators, or as the mode sum (H.c.\ is the
Hermitian conjugate)
\begin{equation}\label{rindlerfield}
 \hat\phi (\vec{x} ,t) =
 \sum_i \left[ w_{Ri}(\vec{x}, t) \, \hat{a}_R( w_{Ri} ) + \mbox{H.c.} 
 \right] \; ,
\end{equation}
in the right Rindler wedge $\{ x > |t| \}$ of Minkowski spacetime
(where the accelerating qubit's worldline is contained),
where $\{ w_{Ri} \}$ are an orthonormal basis of
solutions on the right Rindler wedge which are
positive-frequency with respect to the timelike boost Killing
field $\hat{a} (x \partial_t + t \partial_x )$ there (which is normalized
to have unit length along our qubit's worldline)
and $\hat{a}_R ^{\dagger} (w_{Ri} )$ and $\hat{a}_R (w_{Ri} )$ denote
the corresponding creation and annihilation operators. 

There is a similar mode sum for $\hat\phi (\vec{x},t)$ on the left
Rindler wedge (see~\cite{UnruhWald,wald94} for details). The operators
$\hat{b}^\dagger$ and $\hat{b}$ in Eq.\,(\ref{gwham}) denote the
raising and lowering operators in the internal (two-dimensional)
Hilbert space of the qubit: 
\begin{eqnarray}
 \hat{b} |0\rangle = 0 \; , & \quad &  \hat{b} |1\rangle = |0\rangle\; , \cr
 \hat{b}^\dagger |0\rangle = |1\rangle \; , & \quad & \hat{b}^\dagger
 |1\rangle  = 0 \; . 
\end{eqnarray}
Accordingly, the internal Hamiltonian of the qubit is simply
\begin{equation}
 H_0 = \Omega \, \hat{b}^\dagger \hat{b} \; ,
\end{equation}
where $\Omega$ is the energy difference
between the ground and excited states.

As first discovered by Unruh~\cite{Unruh}, when the (bosonic)
Fock space for the quantum theory
of the field $\phi$ is built from the Rindler mode expansion
Eq.\,(\ref{rindlerfield}) instead of the Minkowski-mode expansion
Eq.\,(\ref{grfield}), the Minkowski vacuum state $|0\rangle_M$ can be
written in the form 
\begin{equation}\label{mv}
 |0\rangle_M = \prod_i \left(
 C_i \sum_{n_i} e^{- \pi n_i \omega_i /a } |n_i , R \rangle
 \otimes |n_i , L \rangle \right) \; .
\end{equation}
Here the product is over all Rindler modes of the form $w_{Ri}$ (and
the corresponding left-wedge modes $w_{Li}$) with (positive) frequencies
$\omega_i$, and the inner sum is over all non-negative integers $n_i$
with $|n_i ,R\rangle$ and $|n_i , L \rangle$ denoting the state with
$n_i$ particles in the mode $w_{Ri}$ and $w_{Li}$ respectively. The
normalization constants $C_i$ are given by
\begin{equation}
 C_i \equiv \sqrt{ 1 - e^{- 2 \pi \omega_i /a } } \; .
\end{equation}
When the left-Rindler components of the vacuum state $|0\rangle_M$
of Eq.\,(\ref{mv}) are traced-out, the reduced density matrix for Minkowski
vacuum as seen by uniformly accelerating observers in the right Rindler
wedge is given by
\begin{equation}\label{thermal}
 \rho_R = \prod_i \left(
 {C_i}^2 \sum_{n_i} e^{- 2 \pi n_i \omega_i /a } |n_i , R \rangle
 \otimes \langle n_i , R | \right) \; 
\end{equation}
which is exactly a thermal state at the Unruh temperature
\begin{equation}
 k_B T = \frac{a}{2 \pi}  \; .
\end{equation}

\subsection{Evolution of the qubit}

We are interested in calculating the final state of the qubit when initially,
before the acceleration commences, it is in the state
\begin{equation}\label{input47}
 |\psi\rangle = \frac{|0 \rangle + | 1 \rangle}{\sqrt{2}} \; .
\end{equation}
Let us first assume that initially
the qubit is in the ground state $|0\rangle$, and the field is in the
Fock state $|n\rangle$ having $n$ particles in the mode
corresponding to a one-particle Hilbert space
element (positive-frequency solution) $\chi (\vec{x}, t)$. So the combined
initial state of the field and the qubit is
\begin{equation}\label{inputstate}
 |\mbox{in}\rangle = |n\rangle \otimes |0\rangle \; .
\end{equation}
The evolution of this input state under the interaction Hamiltonian
Eq.\,(\ref{gwham}) is computed in first order perturbation theory
(in the coupling constant $\epsilon$)
in~\cite{UnruhWald}, between Eqs.\,(3.4) and (3.26)
there. To state their result, let us
introduce the smooth function of compact support
\begin{equation}\label{eq49}
 f(\vec{x}, t) \equiv \epsilon (t) \, e^{i \Omega t} \psi^{\ast} (\vec{x})
\end{equation}
and the retarded minus advanced solution of the Klein-Gordon equation
with source $f$: ${\cal E} f \equiv Rf - Af$, where $\square
(Rf) = \square (Af) = f$. Denote the positive and negative frequency
parts of ${\cal E} f$ by $\Gamma_+$ and $\Gamma_-$, respectively, as
in~\cite{UnruhWald}. We have the basic relation
\begin{eqnarray}
 \hat\phi (f) & \equiv & \int \hat\phi (\vec{x},t)\,
 f(\vec{x},t) \, \sqrt{-g} \, d^4 x \nonumber \\
 & = & \hat{a}(\Gamma^{\ast}_- ) - \hat{a}^\dagger (\Gamma_+ ) \; 
\end{eqnarray}
expressing the field operator smeared with the test function $f$ in
terms of the creation and annihilation operators acting on Fock space.
Then, at the end of the acceleration (when $t \gg \Delta$), the final
state of the field and the qubit system evolving under the interaction
Hamiltonian Eq.\,(\ref{gwham}) is, according to first-order
perturbation theory, 
\begin{equation}\label{out51}
 |\mbox{out}\rangle =
 |n\rangle \otimes |0 \rangle - i \hat{a}(\Gamma^{\ast}_- ) |n\rangle
 \otimes |1\rangle \; .
\end{equation}
Here
\begin{eqnarray}\label{eq52}
 \hat{a}(\Gamma^{\ast}_- ) |n\rangle &=& \sqrt{n} ( \Gamma^{\ast}_- ,
 \chi ) |n-1 \rangle \nonumber \\
 &=& \sqrt{n} \int\!\! f(\vec{x},t) \chi (\vec{x}, t)
 \sqrt{-g} \, d^4 x |n-1\rangle \; ,
\end{eqnarray}
where $(,)$ denotes the inner product on the 1-particle Hilbert space of
positive-frequency solutions. Assume now that the initial state is
\begin{equation}\label{input53}
 |\mbox{in}\rangle = |n\rangle \otimes |1\rangle \; 
\end{equation}
instead of Eq.\,(\ref{inputstate}).
The evolution of this input state can be
computed in exactly the same way as
in~\cite{UnruhWald}, with only slight modifications
of their Eqs.\,(3.18) to (3.25). It
follows that at the end of the acceleration (when $t \gg \Delta$), the final
state evolving from Eq.\,(\ref{input53})
according to first-order perturbation theory is
\begin{equation}\label{out54}
|\mbox{out}\rangle =
|n\rangle \otimes |1 \rangle + i \hat{a}^\dagger
(\Gamma_+ ) |n\rangle \otimes |0\rangle \; ,
\end{equation}
where
\begin{eqnarray}\label{eq55}
& & \hat{a}^\dagger (\Gamma_+ ) |n\rangle = 
\sqrt{n+1} \, \nu \,|\hat{\Gamma}_+ \otimes_S (n \cdot \chi )\rangle \;
, \nonumber \\
& & \nu  \equiv  \sqrt{(\Gamma_+ , \Gamma_+ )} \; , \; \; \; \; \; \;
\hat{\Gamma}_+ \equiv \Gamma_+ / \nu \; .
\end{eqnarray}
Here $|\hat{\Gamma}_+ \otimes_S (n \cdot \chi ) \rangle$ denotes the Fock state
that corresponds to the symmetrized product of the $n+1$
{\em normalized} positive-frequency solutions (one-particle
states): $\hat{\Gamma}_+ \cdot \chi \cdot
\chi  \cdots  \chi$. Putting together Eqs.\,(\ref{out51}) and (\ref{out54}),
when the initial state of the field and the qubit
system is the coherent superposition
\begin{equation}\label{input56}
 |\mbox{in}\rangle = |n\rangle \otimes (|0\rangle
 + | 1 \rangle )/\sqrt{2} \; ,
\end{equation}
the outgoing state at the end of the acceleration/interaction episode is
\begin{eqnarray}\label{output57}
|\mbox{out}\rangle & = &
|n\rangle \otimes (|0\rangle
+ | 1 \rangle )/\sqrt{2} \nonumber \\
&& -  i \sqrt{n}(\Gamma^{\ast}_- , \chi ) |n-1 \rangle \otimes | 1
|\rangle/\sqrt{2}  \nonumber \\
&& +  i \nu \, \sqrt{n+1} \; |\, \hat{\Gamma}_+ \otimes_S
(n \cdot \chi ) \, \rangle\otimes |0 \rangle/\sqrt{2} \; .
\end{eqnarray}
The physical interpretation of Eq.\,(\ref{output57}) is simple: the first term
is the zeroth-order unperturbed initial state, the second term
corresponds to the absorption of a single quantum of $\phi$-radiation by the
qubit, and the third term corresponds to the stimulated emission of a
single quantum at the transition frequency $\Omega$. When the initial
state of the field is a more general Fock state containing $n_1$
particles in mode $\chi_1$, $n_2$ particles in mode $\chi_2$, and so on:
\begin{equation}
 |\mbox{in}\rangle = |n_1 , n_2 , \cdots , n_q
 \rangle \otimes (|0\rangle + | 1 \rangle )/\sqrt{2} \; ,
\end{equation}
it is not difficult to calculate the appropriate
generalization of Eq.\,(\ref{output57}) for the outgoing state; the result is:
\begin{eqnarray}\label{eq59}
 & \; & |\mbox{out}\rangle = 
 |n_1 , n_2 , \cdots , n_q \rangle \otimes (|0\rangle
 + | 1 \rangle )/\sqrt{2} \nonumber \\
 & - &  \frac{i}{\sqrt{2}} \sum_{k=1}^{q}
 \sqrt{n_k}
 (\Gamma^{\ast}_- , \chi_k ) |
 n_1 , \cdots n_k -1 , \cdots , n_q \rangle
 \otimes | 1 \rangle
 \nonumber \\
 & + &  \frac{i \nu}{\sqrt{2}} 
 \sqrt{1+ \sum_{k=1}^{q} n_k }
 \; |\, \hat{\Gamma}_+ \! \! \otimes_S \! (n_1 \cdot \chi_1 )
 \cdots (n_q \cdot \chi_q ) \, \rangle
 \otimes |0 \rangle \; . \nonumber \\
 & ~ & ~~~~~~~
\end{eqnarray}

We now have all the machinery we need to compute the evolution of the
internal state of an accelerating qubit which is initially in
the state Eq.\,(\ref{input47}). Notice that at no point in
the above formalism, Eqs.\,(\ref{inputstate}) to (\ref{eq59}), we made
any reference to the 
particular construction of the Fock space for the field $\phi$. In
particular, the formalism is valid for the Rindler construction based on
the mode sum Eq.\,(\ref{rindlerfield}), which is the natural
construction of the quantum 
field theory of $\phi$ from the viewpoint of an observer accelerating
with the qubit along the same worldline. According to such an observer,
the Minkowski vacuum state in the right Rindler wedge is given by the
thermal density matrix Eq.\,(\ref{thermal}). The initial state of
the combined field and qubit system at the start of the acceleration is
\begin{eqnarray}\label{eq60}
 \rho_{\rm in} &=& \rho_R \otimes |\psi\rangle\langle\psi| \cr
 &=& \frac{1}{2} \rho_R \otimes (|0\rangle + |1\rangle ) \otimes (\langle 0|
 + \langle 1 | )\; .
\end{eqnarray}
We will now make the formal simplification that the thermal state
Eq.~(\ref{thermal}) contains only one Rindler mode $w_{Ri}=\chi_\Omega$,
where $\chi_\Omega$ is a mode whose
frequency $\omega$ equals the transition frequency $\Omega$.
This simplification is justified because the overlap factor
$({\Gamma^{\ast}}_- , \chi )$ in the absorption (second) term in 
Eq.~(\ref{output57})
is negligible unless $\chi$ happens to be
a mode at a frequency $\omega \approx \Omega$ [see
Eq.\,(\ref{eq52})]. While the 
spontaneous emission terms into states with other modes are not
similarly negligible, they have a trivial form whose contribution is
qualitatively the same with or without the simplification. With this
assumption, the product over modes in Eq.\,(\ref{thermal})
disappears, and the thermal 
state of Rindler particles bathing our qubit can be written in the form
\begin{equation}
 \rho_R = {C}^2 \sum_{n} e^{- 2 \pi n \Omega /a } |n , R \rangle
 \otimes \langle n , R |  \; ,
\end{equation}
where
\begin{equation}\label{eq62}
 C \equiv \sqrt{ 1 - e^{- 2 \pi \Omega /a } } \; .
\end{equation}
The input state of the field and the qubit system, Eq.\,(\ref{eq60}), now takes
the simpler form
\begin{eqnarray}\label{eq63}
 \rho_{\rm in} & = &  \frac{{C}^2}{2} \sum_{n} e^{- 2 \pi n \Omega /a }
 \left [ |n , R \rangle \otimes (|0\rangle
 +  | 1 \rangle ) \right] \nonumber \\
 && \quad\otimes \left[ \langle n , R | \otimes (\langle 0 |
 + \langle 1 | ) \right] \; .
\end{eqnarray}
Each of the components in the sum in Eq.\,(\ref{eq62}) undergoes an
evolution as 
in Eq.\,(\ref{eq59}), ending up at the end of the
acceleration in the final state:
\begin{eqnarray}\label{eq64}
 & & |n , R \rangle \otimes \frac{|0\rangle
 + | 1 \rangle }{\sqrt{2}} \longrightarrow 
 \frac{1}{\sqrt{2Q_n}} \Big{[} \; |n, R\rangle (|0\rangle
 + | 1 \rangle ) \nonumber \\
 & - & i \sqrt{n}
 (\Gamma^{\ast}_- , \chi_\Omega ) |n-1 , R
 \rangle | 1 \rangle \cr
& + & i \nu \, \sqrt{n+1} \; |\, \hat{\Gamma}_+ \otimes_S
(n \cdot \chi_\Omega ) \, , R \rangle
\otimes |0 \rangle \, \Big{]} \; , 
\end{eqnarray}
where we used
\begin{equation}
 Q_n \equiv 1 + \frac{n}{2} | (\Gamma^{\ast}_- , \chi_\Omega ) |^2
 + (n+1) \frac{\nu^2}{2} \; .
\end{equation}
We have corrected the normalization error affecting Eqs.\,(\ref{out51})
through (\ref{eq59}) which is symptomatic of any approximation based on
first-order perturbation theory.

The crucial quantity measuring the strength of the interaction is the
overlap $\mu \equiv (\Gamma^{\ast}_- , \chi_\Omega)$, which, according
to Eqs.\,(\ref{eq49}) and (\ref{eq52}), can be calculated as
\begin{eqnarray}\label{mu}
\mu & \approx & \frac{\epsilon \Delta}{\sqrt{2 \Omega}} \int
\psi^\ast (\vec{x}) e^{i \vec{k}_0 \cdot \vec{x}} \, d^3 x
\nonumber \\
& \approx & 
\frac{\epsilon \Delta}{\sqrt{2 \Omega}}
\, e^{i \zeta} \, e^{- \frac{1}{2} \kappa^2 \Omega^2} \; ,
\end{eqnarray}
where $\vec{k}_0$ denotes the wave vector corresponding to the Rindler
mode $\chi_\Omega$ (thus $|\vec{k}_0 | \approx \Omega$, at least for
small accelerations $a$), $\kappa$ is a length scale setting the
spatial range of the interaction (thus $\psi (\vec{x})$ falls off
like a Gaussian distribution
with variance $\kappa^2$), and $\zeta$ is a phase encoding
directional information about $\vec{k}_0$ which is irrelevant to
our considerations. Similarly to Eq.\,(\ref{mu}), the quantity $\nu$
[Eq.\,(\ref{eq55})] can be computed as
\begin{equation}
\nu \approx \frac{\epsilon \, \Delta}
{2 \left( \sqrt{\pi} \kappa^3 \right)^{1/2}}
\;.
\end{equation}

\begin{figure}[t]
  \begin{center}
        \epsfig{file=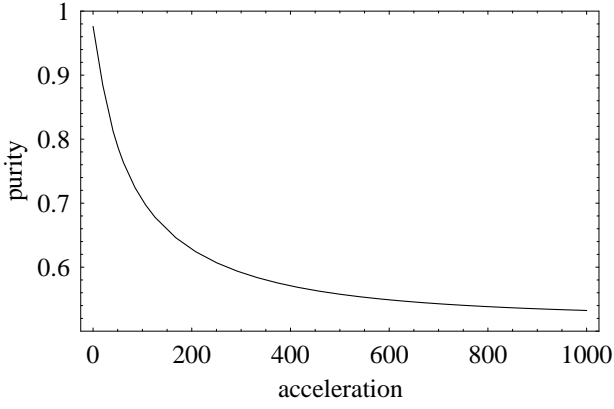,height=5.5cm}
  \end{center}
  \caption{The purity ${\rm Tr}(\rho_{q,\rm out}^2)$ as a function
    of the acceleration of the qubit. The acceleration is plotted
    in units of $\Omega$ (i.e.\ the horizontal axis depicts the
    dimensionless quantity $a/ \Omega$).When there is no acceleration
    the qubit remains pure (purity$=1$), and for increasing
    acceleration the qubit approaches the maximally mixed state
    (purity$=0.5$).} 
  \label{fig:acc}
\end{figure}

Getting back to the evolution of the input state Eq.\,(\ref{eq60}),
combining Eq.\,(\ref{eq64}) with Eq.\,(\ref{eq63}) gives
\begin{eqnarray}\label{eq67}
\rho_{\rm out} & = &  \frac{{C}^2}{2} \frac{\sum_{n} e^{- 2 \pi n \Omega /a }}
{Q_n^{-1}} \Big{[} \; |n , R \rangle \otimes (|0\rangle
+ | 1 \rangle )  \nonumber \\
&& - i \mu \sqrt{n}\, |n-1 , R\rangle \otimes | 1 \rangle\nonumber \\
&& + i \nu \, \sqrt{n+1} \; |\, \hat{\Gamma}_+ \otimes_S
(n \cdot \chi_\Omega ) \, , R \rangle
\otimes |0 \rangle \; \Big{]} \nonumber \\
&& \otimes \Big{[} \; \langle n , R | \otimes (\langle 0 |
+ \langle 1 | ) \nonumber \\
&& +  i \mu^\ast \sqrt{n}
\, \langle n-1 , R
| \otimes \langle 1 |
\nonumber \\
&& - i \nu \, \sqrt{n+1} \; \langle \,
\hat{\Gamma}_+ \otimes_S
(n \cdot \chi_\Omega ) \, , R \, |
\otimes \langle 0 | \; \Big{]} \; .
\end{eqnarray}
All we need to do now to evaluate the final state of the qubit is to
trace over the field degrees of freedom:
\begin{equation}
\rho_{q,\rm out} = \mbox{Tr}_{\phi} \left( \rho_{\rm out}\right) \; .
\end{equation}
It is straightforward to calculate this partial trace of Eq.\,(\ref{eq67}) over
the Fock space; the result is
\begin{eqnarray}\label{eq69}
\rho_{q, \rm out} & = &
 \frac{{C}^2}{2} \sum_{n} \frac{e^{- 2 \pi n \Omega /a }}{Q_n} 
 \left[ (|0\rangle + | 1 \rangle ) (\langle 0 |
 + \langle 1 | ) \right. \cr
 && \left. + |\mu|^2 \, n \, | 1 \rangle \langle 1 | + 
 \nu^2  \, (n+1) \, |0 \rangle \langle 0 | \right] \; ,
\end{eqnarray}
where [cf.\ Eq.\,(\ref{eq64})]
\begin{equation}
Q_n = 1 + n  \frac{| \mu |^2}{2}
+ (n+1)  \frac{\nu^2}{2} \; .
\end{equation}
Introducing the sums
\begin{subequations}
\begin{eqnarray}
 S_0 & \equiv & (1-e^{-2 \pi \Omega /a})
 \sum_{n} \frac{e^{-2 \pi n \Omega /a}}{Q_n} \; , \\
 S_a & \equiv & (1-e^{-2 \pi \Omega /a})  |\mu|^2
 \sum_{n} \frac{n \, e^{-2 \pi n \Omega /a}}{Q_n} \; , \\
 S_e & \equiv & (1-e^{-2 \pi \Omega /a}) \nu^2
 \sum_{n} \frac{(n+1) \, e^{-2 \pi n \Omega /a}}{Q_n} \; ,
\end{eqnarray}
\end{subequations}
which satisfy $S_0 + S_a/2 + S_e/2 = 1$,
Eq.\,(\ref{eq69}) can be written in the
more compact form
\begin{equation}
\rho_{q, \rm out} = \frac{1}{2}
\left(
    \begin{array}{cc}
      S_0 + S_e & S_0 \\
      S_0 & S_0 + S_a
    \end{array}
  \right) \; \; .
\end{equation}

The purity ${\rm Tr}(\rho_{q,\rm out}^2)$ can now be
expressed as a function of the acceleration; a numerical
plot is shown in Fig.~\ref{fig:acc}.

~~~~~~~~~~~

\section{Conclusions}

In conclusion, we have studied the effect of quantum fluctuations in
the spacetime metric on the evolution of a qubit. We considered
three general-relativistic paradigms:
flat fluctuations in a two-dimensional Minkowski
space, mass fluctuations in the Schwarzschild metric for a qubit in a
circular orbit, and the interaction of a qubit with coherent and squeezed
gravitational waves. Finally, we analyzed
the decoherence of an accelerating qubit interacting with the
thermal bath of Unruh particles in Minkowski spacetime.

We found that in addition to the expected decoherence of the qubit, flat
fluctuations induce a phase drift proportional to the square of the
magnitude of the fluctuations. Similarly, mass fluctuations in a black
hole induce a phase shift in addition to the decoherence.
Gravitational radiation has a peculiar
interaction with the qubit, involving periodic episodes of decoherence
followed by restoration of coherence for a sinusoidal wave.
These periodic episodes disappear when we consider wave packets
which, in both the coherent and squeezed states
induce a small phase drift proportional to the strength
of the field as well as the expected decay of off-diagonal
density-matrix elements. The thermal Unruh radiation induces decoherence with a
characteristic dependence on the magnitude of the acceleration.

\section*{Acknowledgements}

The research described in this paper was carried out at the Jet
Propulsion Laboratory, California Institute of Technology, under a
contract with the National Aeronautics and Space Administration
(NASA), and was supported by grants from NASA and the Defense Advanced
Research Projects Agency. P.K. was supported by the United States
National Research Council.  We thank Samuel Braunstein, Gerard Milburn
and Robert Gingrich for helpful discussions.

\end{document}